\documentclass{csmagclass}														

\begin{document}		
\headings																															 %

\title{{ Unsaturated bipartite entanglement of a spin-1/2 Ising-Heisenberg model on a triangulated Husimi lattice}}
\author[1]{{ C. EKIZ\thanks{Corresponding author: cekiz@adu.edu.tr}}}
\author[2]{{ J. STRE\v{C}KA}}
\affil[1]{{Faculty of Science and Art, Ayd{\i}n Adnan Menderes University, 09010 Ayd{\i}n, Turkey}}
\affil[2]{{Faculty of Science, P. J. \v{S}af\'{a}rik University, Park Angelinum 9, 04001 Ko\v{s}ice, Slovakia}}

\maketitle

\begin{Abs}
A bipartite entanglement between two nearest-neighbor Heisenberg spins of a spin-1/2 Ising-Heisenberg model on a triangulated Husimi lattice is quantified using a concurrence. It is shown that the concurrence equals zero in a classical ferromagnetic and a quantum disordered phase, while it becomes sizable though unsaturated in a quantum ferromagnetic phase. A thermally-assisted reentrance of the concurrence is found above a classical ferromagnetic phase, whereas a quantum ferromagnetic phase displays a striking cusp of the concurrence at a critical temperature.
 \end{Abs}
\section{Introduction}
The polymeric compound Cu$_9$Cl$_2$(cpa)$_6$$\cdot$nH$_2$O (cpa=carboxypentonic acid) has recently attracted a lot of attention, because it does not order down to the lowest experimentally reached temperatures due to a geometric spin frustration of the underlying triangulated kagom\'e lattice \cite{ham18,far18}. Because of intractability of the respective spin-1/2 Heisenberg model on a triangulated kagom\'e lattice we have proposed and exactly solved a simpler spin-1/2 Ising-Heisenberg model on related triangulated (triangles-in-triangles) structures with the aim to bring insight into unconventional magnetism of this highly frustrated magnetic material \cite{str08,str15}. From this perspective, the spin-1/2 Ising-Heisenberg model on a triangulated kagom\'e lattice \cite{str08} and its related recursive triangulated Husimi counterpart \cite{str15} affords a long sought-after playground for a theoretical investigation of the quantum entanglement, which is eligible also for an experimental testing.    
\section{Ising-Heisenberg model on a triangulated Husimi lattice}
\begin{figure}
\vspace{-0.5cm}
\begin{center}
\includegraphics[width=0.7\textwidth]{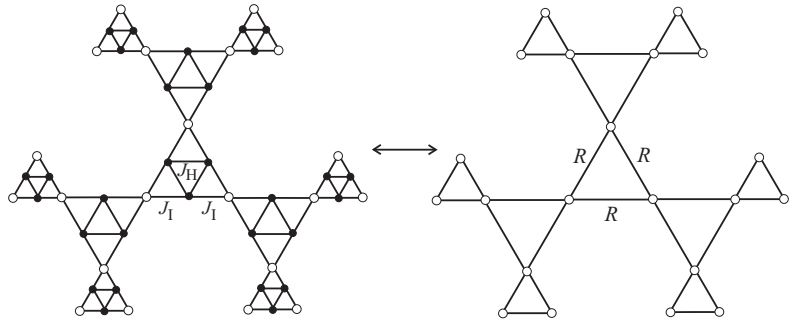}
\end{center}
\vspace{-0.6cm}
\caption{The spin-1/2 Ising-Heisenberg model on a triangulated Husimi lattice (left-hand-side) and its rigorous mapping to the effective spin-1/2 Ising model on a triangular Husimi lattice (right-hand-side). Filled (empty) circles denote lattice positions of the Heisenberg (Ising) spins.}
\label{fig1}
\end{figure}

The spin-1/2 Ising-Heisenberg model on a triangulated Husimi lattice schematically illustrated on the left-hand-side of Fig. \ref{fig1} can be defined through the Hamiltonian
\begin{eqnarray}
\hat{\cal H} = -J_{\rm H} \sum_{\langle k,l \rangle}^{2N} [\Delta(\hat{S}_{k}^{x}\hat{S}_{l}^{x}+\hat{S}_{k}^{y}\hat{S}_{l}^{y})+\hat{S}_{k}^{z}\hat{S}_{l}^{z}]
           -J_{\rm I} \sum_{\langle k,j \rangle}^{4N} \hat{S}_{k}^{z} \hat{\sigma}_{j}^{z}, \nonumber 
\label{eq:1}
\end{eqnarray}
where $\hat{\sigma}^{z}_{j}$ and $\hat{S}_{k}^{\alpha}$ ($\alpha = x,y,z$) label spatial components of the usual spin-1/2 operator assigned to the Ising and Heisenberg spins, respectively, $N$ denotes the total number of the Ising spins, the parameter $J_{\rm H}$ is the XXZ interaction between the nearest-neighbor Heisenberg spins, $\Delta$ is an exchange anisotropy, and the parameter $J_{\rm I}$ labels the Ising interaction between the nearest-neighbor Heisenberg and Ising spins. The overall magnetic structure of the triangulated Husimi lattice form smaller triangles of the Heisenberg spins (trimers), which are embedded into larger triangles of the triangular Husimi lattice involving in its nodal lattice sites the Ising spins. 

It has been previously proved \cite{str15} that the generalized star-triangle transformation provides an exact mapping correspondence between the partition functions and associated free energies of the spin-1/2 Ising-Heisenberg model on a triangulated Husimi lattice and the effective spin-1/2 Ising model on a triangular Husimi lattice shown on right-hand-side of Fig. \ref{fig1}
\begin{eqnarray}
F_{\rm IHM} (\beta, J_{\rm H}, J_{\rm I}, \Delta) = F_{\rm IM} (\beta R) - \frac{2}{3} \ln A.
\label{eq:3}
\end{eqnarray}
Here, $\beta = 1/(k_{\rm B} T)$, $k_{\rm B}$ is Boltzmann's constant, $T$ is absolute temperature and two mapping parameters $A$ and $\beta R$ are given by Eqs. (6)-(9) of Ref. \cite{str15}. The free energy of the effective spin-1/2 Ising model on a triangular Husimi lattice can be found through exact recursive relations
\begin{eqnarray}
F_{\rm IM} = \beta^{-1} [2 \ln ({\rm e}^{\beta R} + 2x + x^2) - \ln(1 + x^2) - \beta R/2]. 
\label{eq:4}
\end{eqnarray}
The parameter $x$ can be obtained by solving the recursive relation (Eq. (13) in Ref. \cite{str15}) iteratively  or by solving the polynomial equation $x^3 + (2 - {\rm e}^{\beta R}) x^2 + ({\rm e}^{\beta R} - 2) x - 1 =0$ with the roots
\begin{eqnarray}
x_{1,2} = \frac{1}{2} \!\left[{\rm e}^{\beta R} - 3 \pm \sqrt{({\rm e}^{\beta R} - 5)({\rm e}^{\beta R} - 1)}\right]\!, \quad x_3=1. 
\label{eq:5}
\end{eqnarray}
It is noteworthy that the first two solutions $x_{1,2}$ correspond to a spontaneously ordered phase with two opposite signs of the spontaneous magnetization, while the third solution $x_3 = 1$ corresponds to a disordered paramagnetic phase without any long-range order. Hence, it follows that the critical temperature of the effective spin-1/2 Ising model on a triangular Husimi lattice is given by the condition $\beta_{\rm c} R = \ln 5$ being consistent with a coalescence of all three roots $x_1=x_2=x_3=1$, which also represents the critical condition for the spin-1/2 Ising-Heisenberg model on a triangulated Husimi lattice due to the mapping relation (\ref{eq:3})  between the free energies. 

The main goal of this work lies in a rigorous analysis of entanglement. A bipartite entanglement between two nearest-neighbor Heisenberg spins can be quantified via concurrence \cite{ami04}
\begin{eqnarray}
C=\max \left \{ 0, 4\left |{\it C_{\rm HH}^{\rm xx}}\right |-2\,\sqrt{\left ( \frac{1}{4}+{\it C_{\rm HH}^{\rm zz}}\right )^{2}-m_{\rm H}^{2}} \right \},
\label{eq:6}
\end{eqnarray}
which can be expressed in terms of three local observables, namely, two spatial components of the pair correlation function 
$C_{\rm HH}^{\rm xx} = \langle \hat{S}_{k,i}^x\hat{S}_{k,i+1}^x \rangle$, $C_{\rm HH}^{\rm zz} = \langle \hat{S}_{k,i}^z\hat{S}_{k,i+1}^z \rangle$ and the sublattice magnetization of the Heisenberg spins $m_{\rm H} = \langle (\hat{S}_{k,i}^z + \hat{S}_{k,i+1}^z)/2 \rangle$. An exact result for the sublattice magnetization $m_{\rm H}$ was already reported in Ref. \cite{str15} [see Eq. (19)], while both spatial components of the pair correlation function can be calculated from Eq. (\ref{eq:3}) according to the formulas
\begin{eqnarray}
C_{\rm HH}^{\rm xx}= -\frac{1}{4} \frac{\partial F_{\rm IHM}}{\partial J_{\rm H} \Delta}, \qquad
C_{\rm HH}^{\rm zz}= -\frac{1}{2} \frac{\partial F_{\rm IHM}}{\partial J_{\rm H}}.
\label{eq:cor}
\end{eqnarray}
The final formulas for $C_{\rm HH}^{\rm xx}$ and $C_{\rm HH}^{\rm zz}$ are too complex to write them down here explicitly.
\section{Results and Discussion}
Let us explore the bipartite entanglement of the spin-1/2 Ising-Heisenberg model on a triangulated Husimi lattice, whereas our further attention will be restricted to the model with the ferromagnetic coupling constant $J_{\rm I} > 0$  because the antiferromagnetic counterpart $J_{\rm I} < 0$ causes a mere flip of all Ising spins. First, we will briefly comment on all possible ground states of the spin-1/2 Ising-Heisenberg model on a triangulated Husimi lattice, which have been already reported in Ref. \cite{str15} and can be classified either as the classical ferromagnetic (CF) phase 
\begin{eqnarray}
\vert \mbox{CF} \rangle = \prod_{i=1}^{N} \left  \vert \uparrow \right \rangle_{\!\sigma_{i}^z} 
                          \prod_{k=1}^{2N/3} \left \vert \uparrow \uparrow \uparrow \right \rangle_{\!S_{k1}^z,S_{k2}^z,S_{k3}^z},
\label{cfp} 
\end{eqnarray}
or the quantum ferromagnetic (QF) phase
\begin{eqnarray}
\vert \mbox{QF} \rangle \!=\! \prod_{i=1}^{N} \! \left \vert \uparrow \right \rangle_{\!\sigma_{i}^z} \!\! \prod_{k=1}^{2N/3} 
\!\! \frac{1}{\sqrt{3}} \! 
\left( \left \vert \uparrow \uparrow \downarrow \right \rangle \!+\! \left \vert \uparrow \downarrow \uparrow \right \rangle 
\!+\! \left \vert \downarrow \uparrow \uparrow \right \rangle \right)_{\!S_{k1}^z,S_{k2}^z,S_{k3}^z}\!\!\!,
\label{qfp} 
\end{eqnarray} 
or the highly degenerate ground-state manifold further referred to as the quantum disordered (QD) phase. The QD phase emerges in the  frustrated parameter space $J_{\rm H}/J_{\rm I}<-2/(2+\Delta)$, the QF phase is realized whenever $\Delta>1$ and $J_{\rm H}/J_{\rm I}>1/(\Delta-1)$, while the CF phase is allocated in the parameter region bounded by the inequalities $J_{\rm H}/J_{\rm I}>-2/(2+\Delta)$ and $J_{\rm H}/J_{\rm I}<1/(\Delta-1)$.

\begin{figure}[t] 
\begin{center}
\vspace*{-0.5cm}
\hspace*{-0.4cm}
\includegraphics[width=0.4\textwidth]{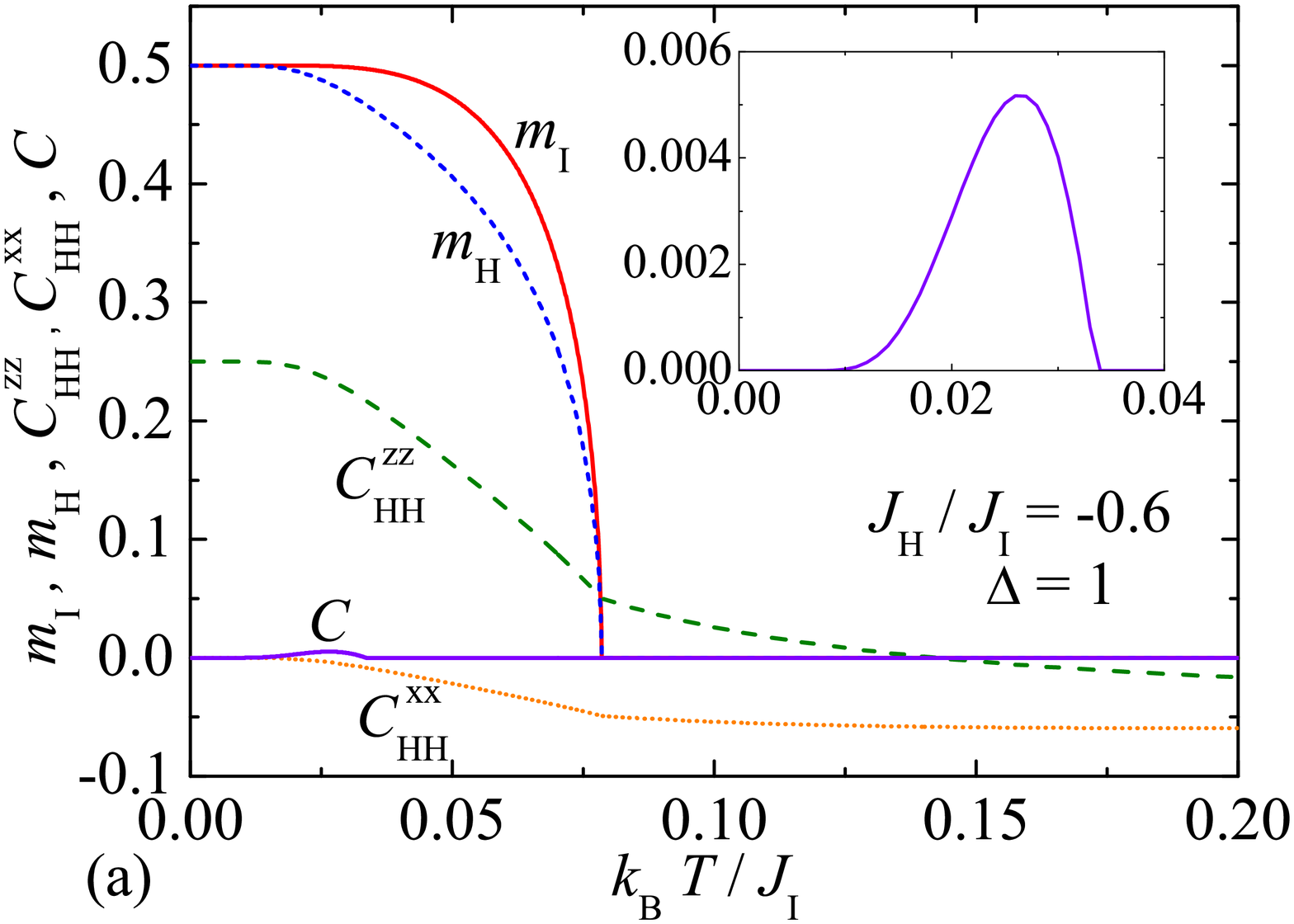}
\hspace*{-0.4cm}
\includegraphics[width=0.4\textwidth]{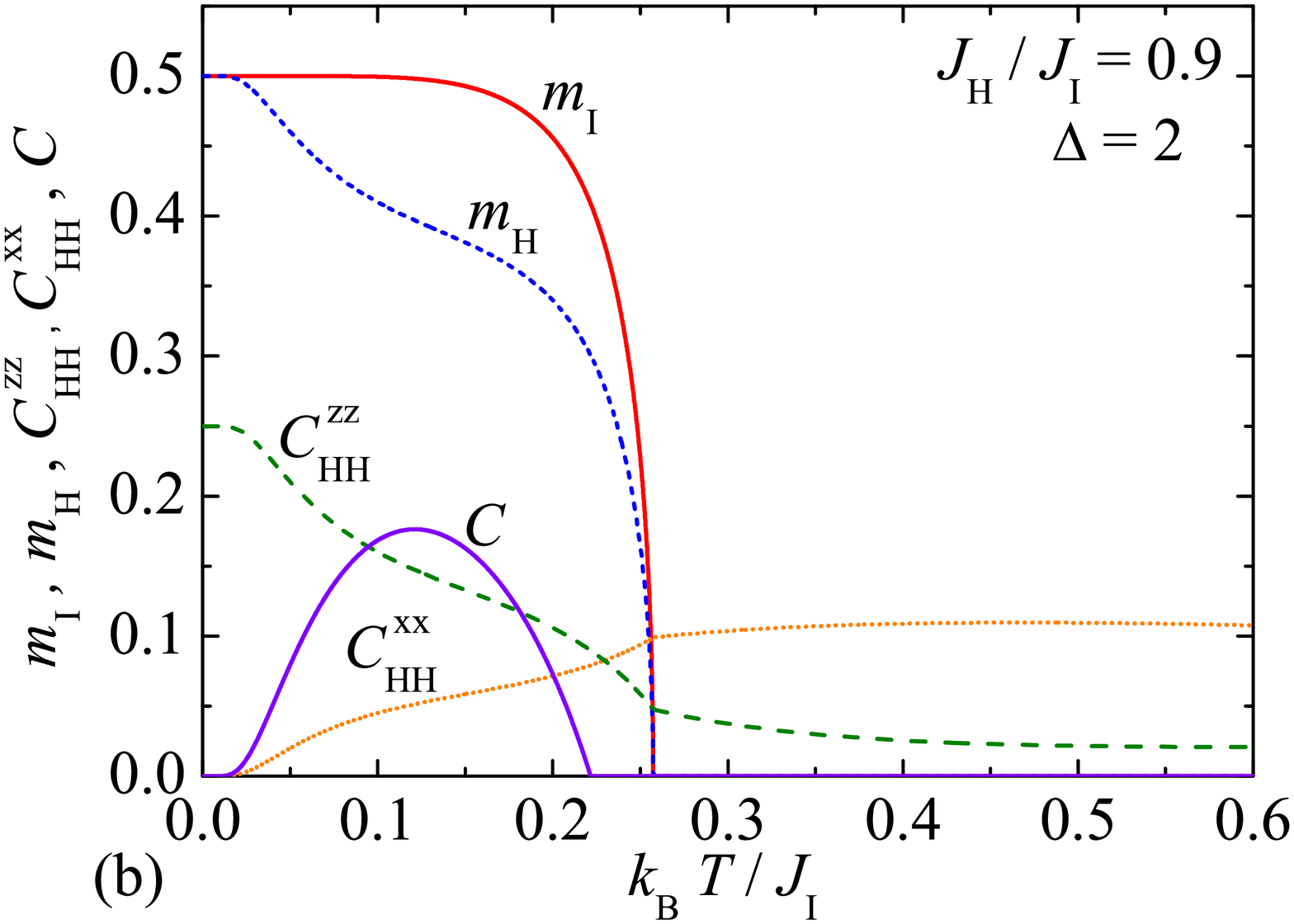}
\hspace*{-0.4cm}
\includegraphics[width=0.4\textwidth]{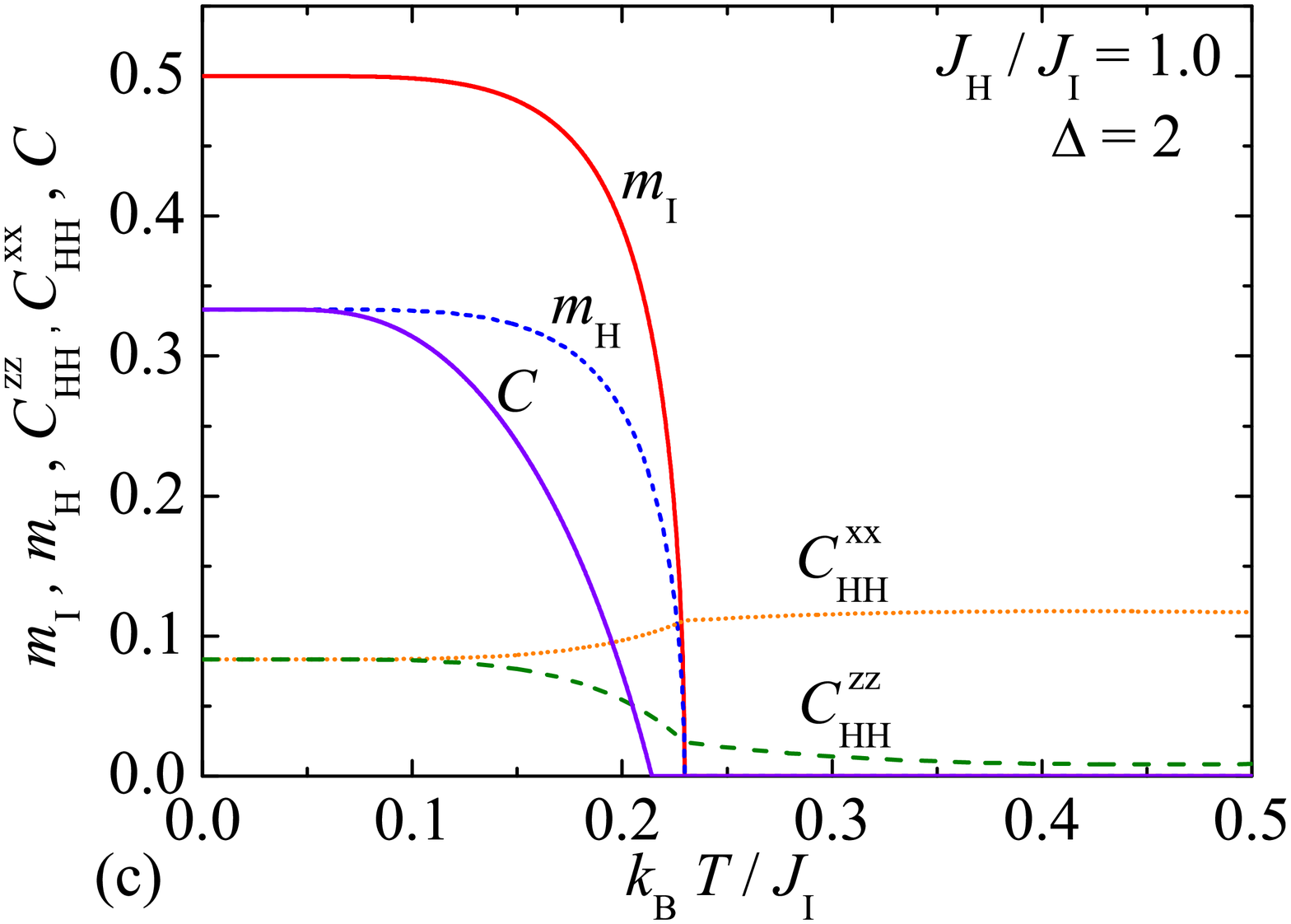}
\hspace*{-0.4cm}
\includegraphics[width=0.4\textwidth]{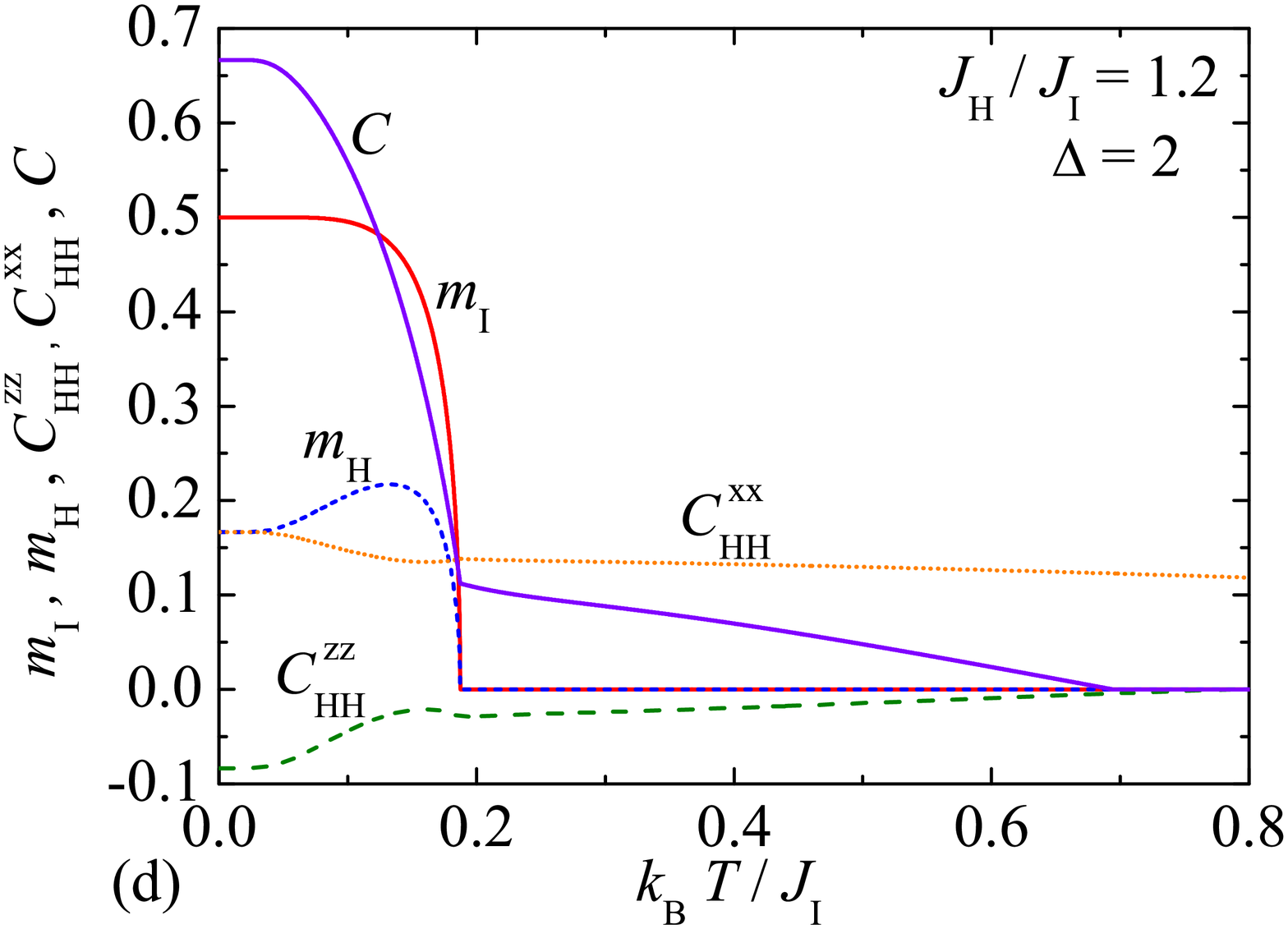}
\vspace*{-0.9cm}
\end{center}
\caption{\small (Color online) Typical thermal variations of the sublattice magnetizations of the Ising ($m_{\rm I}$) and Heisenberg ($m_{\rm H}$) spins, the correlation functions ($C_{\rm HH}^{\rm xx}$, $C_{\rm HH}^{\rm zz}$) and the concurrence $C$ for a few selected sets of the interaction parameters. The inset in Fig. \ref{fig2}(a) shows concurrence in enlarged scale.}
\label{fig2}
\end{figure}

The sublattice magnetizations of the Ising and Heisenberg spins ($m_{\rm I}$, $m_{\rm H}$), the correlation functions ($C_{\rm HH}^{\rm xx}$, $C_{\rm HH}^{\rm zz}$) and the concurrence $C$ are plotted against temperature in Fig. \ref{fig2}(a)-(d). Thermal variations displayed in Fig. \ref{fig2}(a)-(b) show qualitative similarities, since the selected parameters coincide with the CF ground state. It is evident that the concurrence may display a striking thermally-assisted reentrance, which is much more pronounced for the ferromagnetic Heisenberg coupling [$J_{\rm H} > 0$, Fig. \ref{fig2}(b)] than the antiferromagnetic one [$J_{\rm H} < 0$, Fig. \ref{fig2}(a)]. Fig. \ref{fig2}(c) depicts typical behavior at a coexistence point of the CF and QF ground states, while Fig. \ref{fig2}(d) displays typical temperature dependences when starting from the QF ground state. Under this condition, the concurrence decreases upon increasing temperature until it reaches an outstanding cusp at a critical temperature associated with breakdown of the spontaneous long-range order, which is successively followed by a gradual decline ending at a threshold temperature. 

\begin{figure}[t] 
\begin{center}
\vspace*{-0.3cm}
\includegraphics[width=0.45\textwidth]{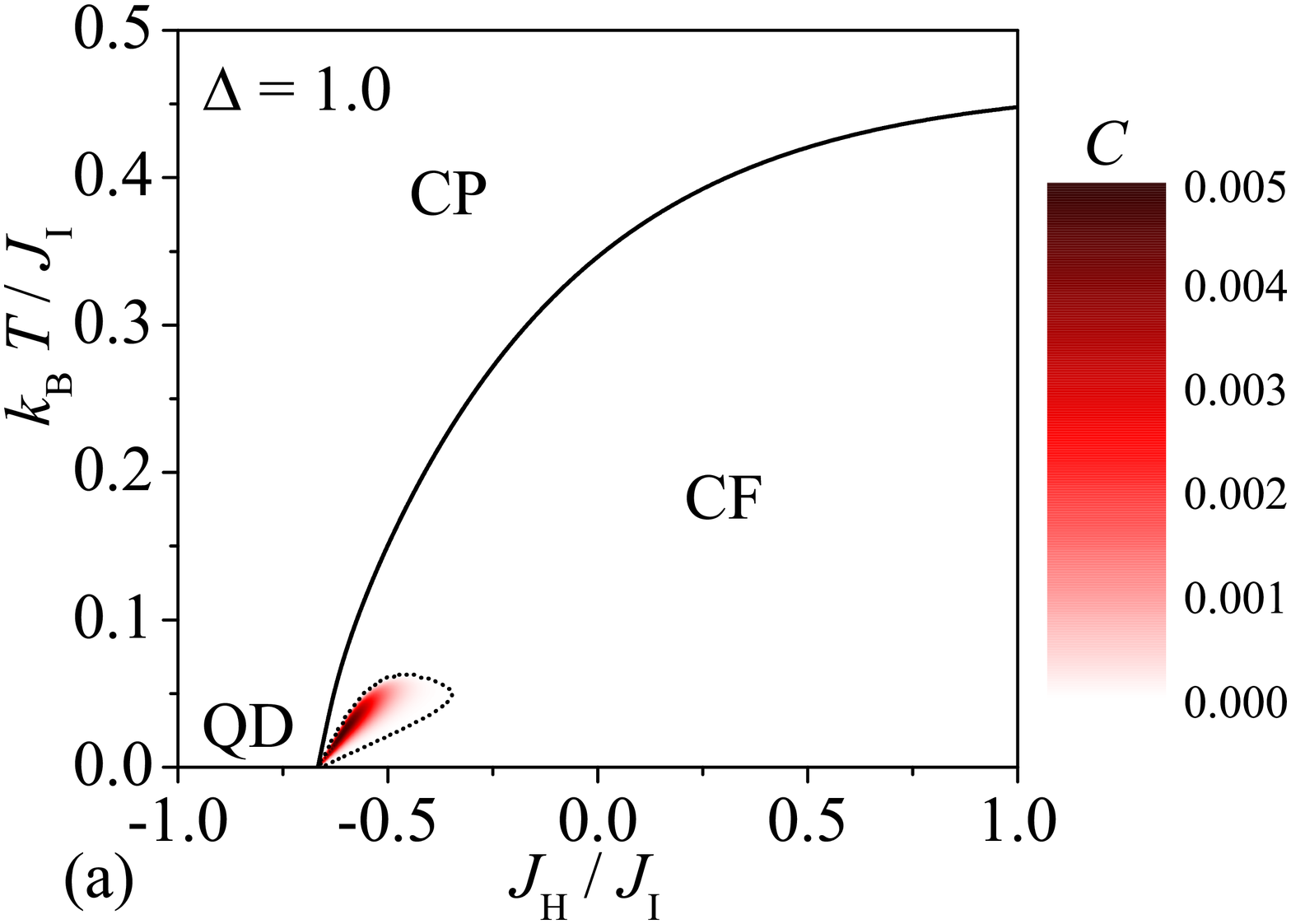}
\vspace*{-0.1cm}
\includegraphics[width=0.45\textwidth]{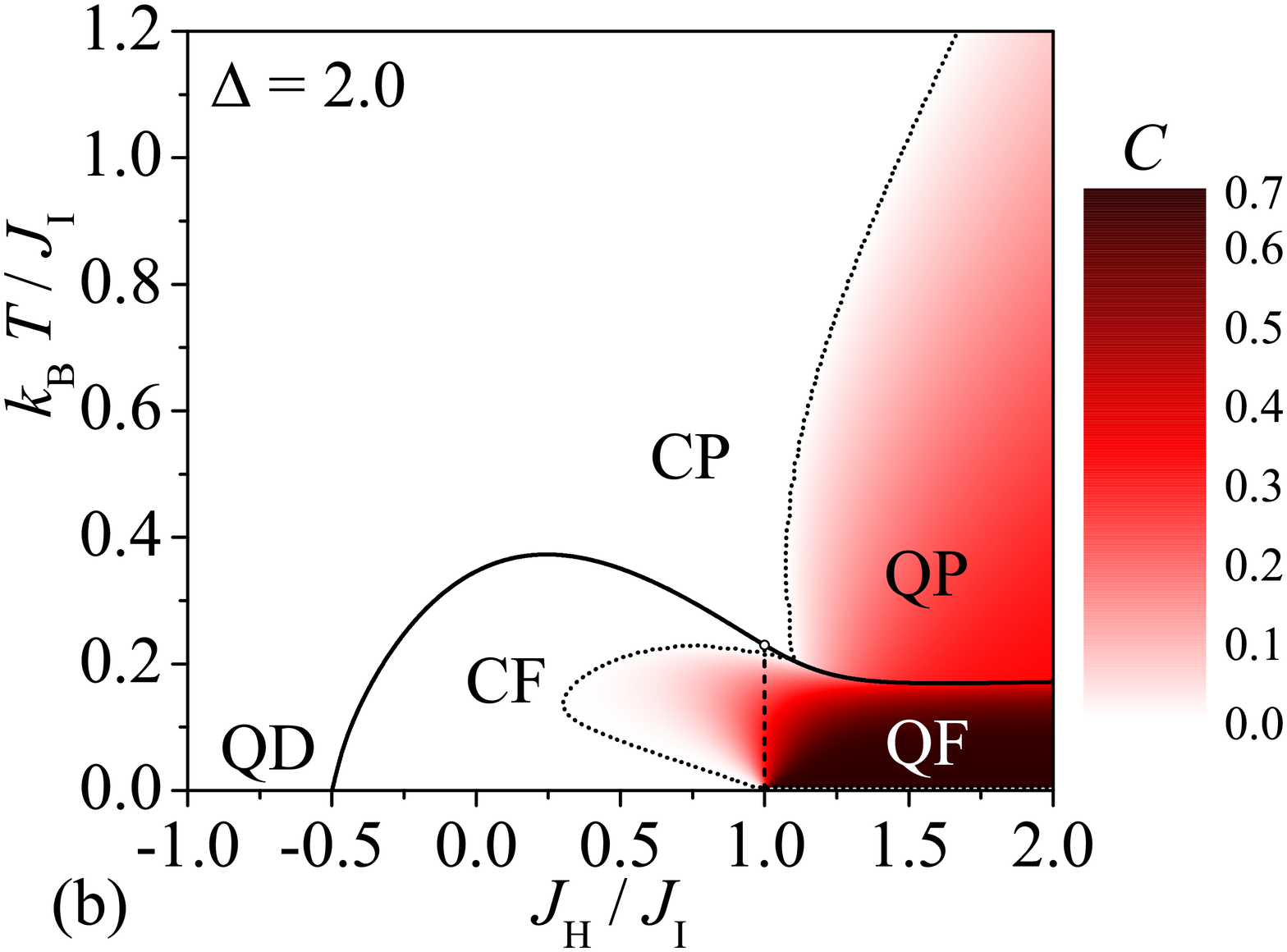}
\vspace*{-0.7cm}
\end{center}
\caption{\small (Color online) A density plot of the concurrence in the $J_{\rm H}/J_{\rm I} - k_{\rm B} T/J_{\rm I}$ plane for two different values of the exchange anisotropy: (a) $\Delta=1$; (b) $\Delta=2$. Dotted lines separate the entangled region ($C > 0$) from the disentangled one ($C = 0$), while solid lines display a critical temperature associated with a breakdown of the spontaneous long-range order.}
\label{fig3}
\end{figure}

Let us summarize our findings by constructing global phase diagrams of the spin-1/2 Ising-Heisenberg model on a triangulated Husimi lattice for two values of the exchange anisotropy $\Delta=1$ and $2$. To this end, a density plot of the concurrence is shown in Fig. \ref{fig3}(a)-(b) in $J_{\rm H}/J_{\rm I} - k_{\rm B} T/J_{\rm I}$ plane along with a critical temperature connected with a breakdown of the spontaneous order. It follows from Fig. \ref{fig3}(a)-(b) that a weak entanglement can be found within the CF phase close to a phase boundary either with the QD or QF phase. However, the preponderant entanglement can be detected in the QF phase, which exhibits a sizable drop of the concurrence around the critical temperature successively followed by a more gradual thermally-assisted decline. The concurrence thus survives far above the critical temperature of the QF phase.
\section{Conclusions}
In the present work, the quantum entanglement of the spin-1/2 Ising-Heisenberg model on a triangulated Husimi lattice has been examined in detail. Exact results for the sublattice magnetization and two spatial components of the pair correlation function were employed in order to calculate the quantum concurrence, which serves as a measure of the bipartite entanglement between the nearest-neighbor Heisenberg spins. It has been found that the bipartite entanglement is totally absent within the CF and QD ground states, while it becomes sizable though unsaturated within the QF ground state. Strikingly, a thermally-assisted reentrance of a relatively weak bipartite entanglement (concurrence) can be detected above the CF ground state in a vicinity of phase boundary either with the QF or QD ground states. In addition, it turns out that the threshold temperature, above which the bipartite entanglement  vanishes, may seemingly exceed the critical temperature of the QF phase accompanied with a cusp in the relevant temperature dependence of the concurrence. 
\section{Acknowledgement}
This work was supported under the grant Nos. VEGA 1/0531/19 and APVV-16-0186.
%


\end{document}